\documentclass[final,10pt]{elsart}

\usepackage{graphicx}

\usepackage{amssymb}

\begin{document}

\begin{frontmatter}



\title{Fractional phase transition in medium size metal clusters and some remarks on magic numbers in gravitationally
and weakly interacting clusters}


\author{Richard Herrmann}
\ead{herrmann@gigahedron.com}

\address{GigaHedron, Berliner Ring 80, D-63303 Dreieich, Germany}

\begin{abstract}
Based on the Riemann- and Caputo definition of the fractional derivative we use the fractional
extensions of the standard rotation group $SO(3)$ to construct a higher dimensional representation
of a fractional rotation group with mixed derivative types. 
An analytic extended symmetric rotor model is derived, which correctly predicts 
the sequence of magic numbers in metal clusters.
It is demonstrated, that experimental data may be described assuming a sudden change in the fractional 
derivative parameter $\alpha$ which is interpreted as a second order phase transition in the region of cluster size
with 
$200 \leq N \leq 300$. 

Furthermore it is demonstrated, that the four different realizations of higher
dimensional fractional rotation groups may successfully be connected to the four fundamental interaction types 
realized in 
nature and may be therefore used for a prediction of magic numbers and binding energies of clusters with
gravitational force and weak force respectively bound constituents. 

The results presented lead to the conclusion, that mixed fractional derivative operators might play 
a key role for a successful unified theoretical description  of all four fundamental forces realized in nature.    
\end{abstract}

\end{frontmatter}

\section{Introduction}
Since 1984, an increasing amount of experimental data{\cite{kni}}-{\cite{mar}} confirms an at first 
unexspected shell structure in fermion systems, realized as magic numbers in metal clusters.

The observation of varying binding energy of the valence electron, moving freely in a metallic cluster, has 
initiated the developement of several theoretical models. Besides ab initio calculation, the most prominent 
representatives are the 
jellium model{\cite{brack}}
and,  in analogy to methods already in use in nuclear physics, phenomenological shell models with modified
potential terms like the Clemenger-Nilsson model or deformed Woods-Saxon potential{\cite{deHeer}}-{\cite{mor}} .  

Although these models describe the experimental data with reasonable accuracy, they do not give a theoretical
explanation for the observed sequence of magic numbers. Therefore the problem 
of a theoretical foundation of the magic numbers is still an open question.

A more fundamental understanding of magic numbers found for metal clusters
may be achieved if the underlying corresponding symmetry of the fermion many body system is determined. 
Therefore a group theoretical approach seems appropriate.
 
In this paper we will determine the symmetry group, which generates the single particle spectrum
of a metal cluster 
similar to jellium or phenomenological shell models, but includes the magic numbers  right from the beginning.

Our approach is based on group theoretical methods developed
within the framework of  fractional calculus.

The fractional calculus \cite{f3}-\cite{he08} provides a set of axioms and methods to extend the coordinate 
and corresponding
derivative definitions in a reasonable way from integer order n to arbitrary order $\alpha$:
\begin{equation}
\{ x^n, {\partial^n \over \partial x^n} \} 
\rightarrow
\{ x^\alpha, {\partial^\alpha \over \partial x^\alpha} \}
\end{equation}
The concept of fractional calculus has attracted mathematicians since the days of Leibniz. In physics,
early applications were studies on non local dynamics, e.g. anomalous diffusion 
or fractional Brownian motion \cite{o1},\cite{pod}.

In the last decade, remarkable progress has been made in the theory of fractional wave equations. 
In 2002, Laskin \cite{las} has derived a fractional Schr\"odinger equation.
Raspini \cite{ras}
has proposed a fractional $\alpha=2/3$ Dirac equation and has shown, that the corresponding $\gamma$ matrices obey
an extended Clifford algebra, which is directly related  to SU(3)-symmetry. We \cite{he05} have proposed an 
extension of the ordinary rotation group SO(n) to the fractional case. Based on this 
fractional rotation group,  an extended fractional symmetric rotor model \cite{he07} was presented, which was used for 
a successful description of the 
ground state band spectra of even-even nuclei. 

The definition of the fractional derivative is not unique, several definitions coexist{\cite{f1}}-{\cite{grun}}.
Recently {\cite{he08b}}, the 
properties of higher dimensional rotation groups  
with mixed Caputo and Riemann type definition of the fractional derivative have been investigated. 
From the four different possible realizations of this 9-dimen\-sional fractional rotation group it has been 
demonstrated, that  the Caputo-Riemann-Riemann decomposition establishes 
a fundamental dynamic symmetry, 
which determines the magic numbers for protons and neutrons  and furthermore 
 describes the ground state properties like binding energies and ground state quadrupole deformations 
of nuclei with reasonable accuracy.  

On the basis of this encouraging result, in the present work we will demontrate, that the Caputo-Caputo-Riemann
decomposition of the 9-dimensional fractional rotation group generates a dynamic symmetry group, which 
determines the magic numbers in metal clusters accurately. Furthermore a comparison with experimental data will
lead to the conclusion, that a fractional phase transition occures at $N>198$. 

Finally we will present arguments, that the four different possible decompositions of the 9-dimensional 
fractional rotation group correspond to the four different interaction types realized in nature and will give
resaonable predictions on the sequence of magic numbers in gravitationally and weakly bound clusters.  
\section{Notation}
We will investigate the spectrum of multi dimensional fractional rotation groups for two different definitions
of the fractional derivative, namely the Riemann- and Caputo fractional derivative. Both types are strongly
related.

Starting with the definition of the fractional Riemann integral 
\begin{equation}
{_\textrm{\tiny{R}}}I^\alpha \, f(x) = 
\cases{
({_\textrm{\tiny{R}}}I_{+}^\alpha f)(x) =  
\frac{1}{\Gamma(\alpha)}   
     \int_{0}^x  d\xi \, (x-\xi)^{\alpha-1} f(\xi)&\qquad \, $x \geq 0$ \cr \\
   ({_\textrm{\tiny{R}}}I_{-}^\alpha f)(x) =  
\frac{1}{\Gamma(\alpha)}   
     \int_x^0  d\xi \, (\xi-x)^{\alpha-1} f(\xi)&\qquad \,   $x<0$ \cr 
}
\end{equation} 
where $\Gamma(z)$ denotes the Euler $\Gamma$-function, 
the fractional Riemann derivative is defined as the result of a fractional integration followed by an
ordinary differentiation:
\begin{equation}
\label{dr}
{_\textrm{\tiny{R}}}\partial_x^\alpha =  \frac{\partial}{\partial x}  {_\textrm{\tiny{R}}}I^{1-\alpha}
\end{equation} 
It is explicitly given by:
\begin{equation}
\label{driemann}
{_\textrm{\tiny{R}}}\partial_x^\alpha \, f(x) = 
\cases{
({_\textrm{\tiny{R}}}\partial_{+}^\alpha f)(x) =  
\frac{1}{\Gamma(1 -\alpha)} \frac{\partial}{\partial x}  
     \int_{0}^x  d\xi \, (x-\xi)^{-\alpha} f(\xi)&$x \geq 0$ \cr \\
   ({_\textrm{\tiny{R}}}\partial_{-}^\alpha f)(x) =  
\frac{1}{\Gamma(1 -\alpha)} \frac{\partial}{\partial x}  
     \int_x^0  d\xi \, (\xi-x)^{-\alpha} f(\xi)&$x<0$ \cr 
}
\end{equation} 
The Caputo definition of a fractional derivative follows an inverted sequence of operations (\ref{dr}).
An ordinary differentiation is followed by a fractional integration
\begin{equation}
{_\textrm{\tiny{C}}}\partial_x^\alpha =   {_\textrm{\tiny{R}}}I^{1-\alpha} \frac{\partial}{\partial x} 
\end{equation} 
This results in:
\begin{equation}
{_\textrm{\tiny{C}}}\partial_x^\alpha \, f(x) = 
\cases{
({_\textrm{\tiny{C}}}\partial_{+}^\alpha f)(x) =  
\frac{1}{\Gamma(1 -\alpha)}   
     \int_{0}^x  d\xi \, (x-\xi)^{-\alpha} \frac{\partial}{\partial \xi}f(\xi)&$x \geq 0$ \cr \\
   ({_\textrm{\tiny{C}}}\partial_{-}^\alpha f)(x) =  
\frac{1}{\Gamma(1 -\alpha)}  
     \int_x^0  d\xi \, (\xi-x)^{-\alpha} \frac{\partial}{\partial \xi}f(\xi)&$x<0$ \cr 
}
\end{equation} 
Applied to  a function set $f(x)=x^{n \alpha}$ using the Riemann fractional derivative definition (\ref{driemann}) we
obtain:
\begin{eqnarray}
{_\textrm{\tiny{R}}}\partial_x^\alpha \, x^{n \alpha}  &=& \frac{\Gamma(1+n \alpha)}{\Gamma(1+(n-1)\alpha)} \, x^{(n-1)\alpha}\\
\label{rx}
&=& {_\textrm{\tiny{R}}}[n]  \, x^{(n-1)\alpha}
\end{eqnarray} 
where we have introduced the abbreviation ${_\textrm{\tiny{R}}}[n]$.

For the Caputo definition of the fractional derivative it follows for the same function set:
\begin{eqnarray}
{_\textrm{\tiny{C}}}\partial_x^{\alpha} \, x^{n \alpha} &=& 
\cases{
\frac{\Gamma(1+n \alpha)}{\Gamma(1+(n-1)\alpha)} \, x^{(n-1)\alpha}&$n > 0$ \cr \nonumber \\
0&$n=0$ \cr
}\cr \\
\label{cx}
&=& {_\textrm{\tiny{C}}}[n]  \, x^{(n-1)\alpha}
\end{eqnarray} 
where we have introduced the abbreviation $ {_\textrm{\tiny{C}}}[n]$. 

Both derivative definitions only differ in the case $n=0$: 
\begin{eqnarray}
{_\textrm{\tiny{C}}}[n]  &=&   {_\textrm{\tiny{R}}}[n] -\delta_{n0}\, {_\textrm{\tiny{R}}}[0]   \\
&=&  {_\textrm{\tiny{R}}}[n] - \delta_{n0}\,  \frac{1}{\Gamma(1-\alpha)}
\end{eqnarray} 
where $\delta_{mn}$ denotes the Kronecker-$\delta$.
We will rewrite equations (\ref{rx}) and (\ref{cx}) simultaneously, introducing the short hand notation 
\begin{equation}
{_\textrm{\tiny{R,C}}}\partial_x^\alpha \, x^{n \alpha}  ={_\textrm{\tiny{R,C}}}[n]  \, x^{(n-1)\alpha}\\
\end{equation}
We now introduce the fractional angular momentum operators or generators of infinitesimal rotations 
in the $i,j$ plane on the $N$-dimensional Euclidean space:
 \begin{equation}
{_\textrm{\tiny{R,C}}}L_{ij}(\alpha)  =
i \hbar(x_i^\alpha {_\textrm{\tiny{R,C}}}\partial_j^\alpha-x_j^\alpha{_\textrm{\tiny{R,C}}}\partial_i^\alpha)
\end{equation}
which result from canonical quantization of the classical angular momentum definition (for details see \cite{he07}).
The commutation relations of the fractional angular momentum operators are isomorphic to the fractional 
extension of the rotational group $SO(N)$
\begin{eqnarray}
{_\textrm{\tiny{R,C}}} \, [L_{ij}(\alpha),L_{kl}(\alpha)] &=& i\hbar
{_\textrm{\tiny{R,C}}}{f_{ijkl}}^{mn}(\alpha){_\textrm{\tiny{R,C}}}L_{mn}(\alpha) \\
& &  \qquad\qquad\qquad i,j,k,l,m,n=1,2,..,N \nonumber
\end{eqnarray}
with structure coefficients ${_\textrm{\tiny{R,C}}}{f_{ijkl}}^{mn}(\alpha)$. Their explicit form depends on the 
function set the fractional
angular momentum operators act on and on the fractional derivative type used. 

The structure coefficients and Casimir-operators of the fractional 
rotation group ${_\textrm{\tiny{R}}}SO^\alpha(3)$ based on the
Riemann fractional derivative definition have been derived in {\cite{he07}} and 
for ${_\textrm{\tiny{C}}}SO^\alpha(3)$ based on the
Caputo fractional derivative definition are given in {\cite{he05}}. We summarize the major results:
\begin{figure}
\begin{center}
\includegraphics[width=80mm,height=61mm]{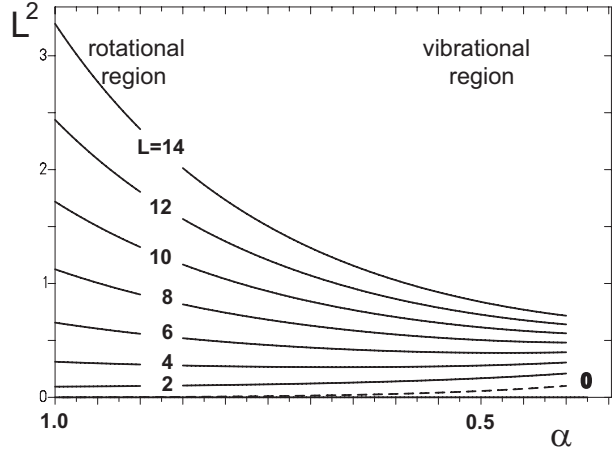}\\
\caption{\label{so3}
Spectrum of the Casimir operator $L^2(L,\alpha)$ from (\ref{L2}) as a function of the fractional derivative coefficient $\alpha$. Only
the $L=0$ state differs for Riemann and Caputo derivative.
} 
\end{center}
\end{figure}

According to the group chain
\begin{equation}
{_\textrm{\tiny{R,C}}}SO^\alpha(3) \supset {_\textrm{\tiny{R,C}}}SO^\alpha(2)
\end{equation}
there are two Casimir-operators $\Lambda_i$, namely $\Lambda_2 =L_z(\alpha)  = L_{12}(\alpha)$ and
$\Lambda_3  = L^2(\alpha) =  L_{12}^2 (\alpha) + L_{13}^2 (\alpha) + L_{23}^2 (\alpha)$. We 
introduce the two quantum numbers $L$ and $M$, which completely determine the eigenfunctions
$|LM>$.
It follows
\begin{eqnarray}
\label{Lz}
{_\textrm{\tiny{R,C}}} \, L_z(\alpha) |LM>  &=&
 \hbar \, \textrm{sign}(M)\, {_\textrm{\tiny{R,C}}} \,[|M|]\, |LM> \\ 
 & & \qquad\qquad\qquad M = -L,-L+1,...,\pm 0,...,L \nonumber \\
\label{L2}
{_\textrm{\tiny{R,C}}} \, L^2(\alpha) |LM>  &=&
{_\textrm{\tiny{R,C}}} \, \hbar^2 {_\textrm{\tiny{R,C}}} \,[L]{_\textrm{\tiny{R,C}}} \,[L+1] \,|LM> \\
 & & \qquad\qquad\qquad  L = 0,1,2,... \nonumber
\end{eqnarray}
where $|M|$ denotes the absolute value of $M$.
In addition, on the set of eigenfunctions $|LM>$, the parity operator $\Pi$ is diagonal and has the
eigenvalues
\begin{equation} 
\label{parity}
\Pi |LM> = (-1)^L |LM>
\end{equation} 
In figure \ref{so3} the eigenvalues of the Casimir-operator $L^2$ are shown as a function of the fractional 
derivative coefficient $\alpha$. As has been pointed out in {\cite{he07}}, near $\alpha \approx 1$ there is a region 
of rotational type of spectrum, while for $\alpha \approx 1/2$,
the levels are nearly equidistant, which corresponds to 
a vibrational type of spectrum.   

In addition, for decreasing $\alpha<1$ higher angular momenta are lowered. 

Only in the case $L=0$ the spectra differ for the Riemann- and Caputo derivative.
While for the Caputo derivative 
\begin{equation}
{_\textrm{\tiny{C}}} \, L^2(\alpha) |00> = 0   
\end{equation}
because
${_\textrm{\tiny{C}}} [0] = 0 $, using the Riemann derivative for $\alpha \neq 1$ there is a non vanishing contribution
\begin{equation} 
\label{gap}
{_\textrm{\tiny{R}}} L^2(\alpha)|00> = \hbar^2 \, {_\textrm{\tiny{R}}} [0]{_\textrm{\tiny{R}}} [1] |00> 
= \hbar^2 \frac{\Gamma(1+\alpha)}{\Gamma(1-\alpha)}|00>
\end{equation}

\section{The Caputo-Caputo-Riemann symmetric rotor}
We now use group theoretical methods to construct higher dimensional representations of the fractional 
rotation groups ${_\textrm{\tiny{R,C}}}SO^\alpha(3)$. 

For a 9-dimensional rotation group $G$ there exist four different decompositions with the following 
chain of sub algebras: 
\begin{eqnarray}
\label{g1}
{_\textrm{\tiny{RRR}}}G &\supset& {_\textrm{\tiny{R}}}SO^\alpha(3)
                        \supset {_\textrm{\tiny{R}}}SO^\alpha(3)
                        \supset {_\textrm{\tiny{R}}}SO^\alpha(3)\\
{_\textrm{\tiny{CRR}}}G &\supset& {_\textrm{\tiny{C}}}SO^\alpha(3)
                        \supset {_\textrm{\tiny{R}}}SO^\alpha(3)
                        \supset {_\textrm{\tiny{R}}}SO^\alpha(3)\\
{_\textrm{\tiny{CCR}}}G &\supset& {_\textrm{\tiny{C}}}SO^\alpha(3)
                        \supset {_\textrm{\tiny{C}}}SO^\alpha(3)
                        \supset {_\textrm{\tiny{R}}}SO^\alpha(3)\\
\label{g4}
{_\textrm{\tiny{CCC}}}G &\supset& {_\textrm{\tiny{C}}}SO^\alpha(3)
                        \supset {_\textrm{\tiny{C}}}SO^\alpha(3)
                        \supset {_\textrm{\tiny{C}}}SO^\alpha(3)
\end{eqnarray}
As an example of physical relevance in {\cite{he08b}} the properties of the 9 dimensional fractional rotation 
group ${_\textrm{\tiny{CRR}}}G$ have already been investigated. The main result was the verification of magic
numbers for nuclei:
\begin{eqnarray} 
\label{setNilsson1}
n_{\textrm{nuclear 1}} &=& \frac{1}{3}(N+1)(N+2)(N+3) \qquad N=0,1,2,3,...\\
&=& 2,8,20,40,70,112,... \\
\label{setNilsson2}
n_{\textrm{nuclear 2}} &=& \frac{1}{3}(N+1)(N+2)(N+3) +  2(N+2) \\
&=& 6,14,28,50,82,126,184,258,...
\end{eqnarray} 
In this section we will discuss the properties of the group ${_\textrm{\tiny{CCR}}}G $:

For that purpose, we associate a Hamiltonian $H$, which can now be written in terms of the Casimir operators of the algebras 
appearing in the
chain and can be analytically diagonalized in the corresponding basis. The Hamiltonian is explicitly given as:
\begin{equation}
\label{hamilton}
H = \frac{\omega_1}{\hbar} {_\textrm{\tiny{C}}} L_1^2(\alpha)+
    \frac{\omega_2}{\hbar} {_\textrm{\tiny{C}}} L_2^2(\alpha)+
    \frac{\omega_3}{\hbar} {_\textrm{\tiny{R}}} L_3^2(\alpha) 
\end{equation}
with the free parameters $\omega_1,\omega_2,\omega_3$ and the basis is $|L_1 M_1 L_2 M_2 L_3 M_3>$.
Furthermore, we demand the following symmetries: 

First, the wave functions should be invariant under
parity transformations, which according to (\ref{parity}) leads to the conditions
\begin{equation}
\label{sym1}
L_1 = 2 n_1 \quad L_2 = 2 n_2 \quad L_3 = 2 n_3, \quad n_1,n_2,n_3=0,1,2,3,...
\end{equation}
second, we require
\begin{eqnarray}
\label{sym2}
{_\textrm{\tiny{C}}}L_{z_1}(\alpha)|L_1 M_1 L_2 M_2 L_3 M_3> &=& +\hbar{_\textrm{\tiny{C}}}[L_1]|L_1 M_1 L_2 M_2 L_3 M_3> \\
{_\textrm{\tiny{C}}}L_{z_2}(\alpha)|L_1 M_1 L_2 M_2 L_3 M_3> &=& +\hbar{_\textrm{\tiny{C}}}[L_2]|L_1 M_1 L_2 M_2 L_3 M_3> \\
{_\textrm{\tiny{R}}}L_{z_3}(\alpha)|L_1 M_1 L_2 M_2 L_3 M_3> &=& +\hbar{_\textrm{\tiny{R}}}[L_3]|L_1 M_1 L_2 M_2 L_3 M_3> 
\end{eqnarray}
which leads to the conditions
\begin{equation}
\label{sym2}
M_1 = 2 n_1 \quad M_2 = 2 n_2 \quad M_3 = 2 n_3, \quad n_1,n_2,n_3=0,1,2,3,...
\end{equation}
and reduces the multiplicity of a given $|2n_1 M_1 2n_2 M_2 2n_3 M_3>$ set to 1.

With these conditions, the eigenvalues of the Hamiltonian (\ref{hamilton}) are given as
\begin{eqnarray}
\label{e1}
E(\alpha) &=& \hbar \omega_1 \, {_\textrm{\tiny{C}}} [2n_1]{_\textrm{\tiny{C}}}[2n_1+1]+ 
      \hbar \omega_2 \, {_\textrm{\tiny{C}}} [2n_2]{_\textrm{\tiny{C}}}[2n_2+1]+\nonumber \\
& & 
      \hbar \omega_3 \, {_\textrm{\tiny{R}}} [2n_3]{_\textrm{\tiny{R}}}[2n_3+1] \\
\label{e2}
&=& \sum_{i=1}^3 \hbar \omega_i \frac{\Gamma(1 + (2 n_i +1) \alpha)}{\Gamma(1 + (2 n_i-1) \alpha)} \nonumber \\
& & 
- \delta_{n_1 0} \hbar \omega_1 \frac{\Gamma(1 +  \alpha)}{\Gamma(1 - \alpha)}
- \delta_{n_2 0} \hbar \omega_2 \frac{\Gamma(1 +  \alpha)}{\Gamma(1 - \alpha)}\\
& &  \qquad\qquad\qquad\qquad\qquad n_1,n_2,n_3=0,1,2,.. \nonumber
\end{eqnarray}
on a basis $| 2 n_1 2 n_1 2 n_2 2 n_2 2 n_3 2 n_3>$.

This is the major result of our derivation. We call this model the Caputo-Caputo-Riemann symmetric rotor. 
What makes this model remarkable is its behaviour in the vibrational region near the semi-derivative $\alpha=1/2$. 

On the left of figure \ref{fig2} we have plotted the energy levels in the vicinity of $\alpha \approx 1/2$
for the case 
\begin{equation}
\label{sphere}
\omega_1=\omega_2=\omega_3=\omega_0
\end{equation}
which we denote as the spherical case. 

For the idealized spherical case $\alpha=1/2$, using the relation $\Gamma(1+z)= z\Gamma(z)$ 
the level spectrum (\ref{e2}) is simply given by:
\begin{equation}
\label{e12}
E(\alpha=1/2) = \hbar \omega_0 (  n_1 +n_2 +n_3 +\frac{3}{2} - \frac{1}{2}\delta_{n_1 0}- \frac{1}{2}\delta_{n_2 0})  
\end{equation}
For $n_1 \neq 0$ and $n_2 \neq 0$ this is the well known spectrum of the 3-dimensional harmonic oscillator. Assuming a twofold
spin degeneracy of the energy levels, we introduce the quantum number $N$ as
\begin{equation} 
N = n_1 + n_2 + n_3 
\end{equation} 
Consequently we obtain the multiplets of the 3-dimensional harmonic oscillator with magic numbers for
\begin{eqnarray} 
\label{setho}
n_{\textrm{\tiny{HO}}} &=& \frac{1}{3}(N+1)(N+2)(N+3) \qquad N=0,1,2,3,...\\
&=& 2,8,20,40,70,112,168,240,...
\end{eqnarray} 
at energies
\begin{equation} 
\label{efirst}
E(N)_{\textrm{\tiny{HO}}} = \hbar \omega_0 (N+3/2)
\end{equation} 
In order to determine the multiplets of (\ref{e12}), we distinguish two different sets of magic numbers:

For $n_1 = 0$ and $n_2 = 0$ 
the multiplicity of a harmonic oscillator shell for $N$ at energy (\ref{efirst}) 
is increased by exactly one state, the $|0000\, N+1 \, N+1>$ state, which originates from the $N+1$ shell. 
Therefore we obtain a first set ${_{\textrm{\tiny{CCR}}}}n_{\textrm{magic 1}} $ of  magic numbers (including the state
$|000000>$): 
\begin{eqnarray} 
\label{set1ccr}
{_{\textrm{\tiny{CCR}}}}n_{\textrm{magic 1}} &=&  n_{\textrm{\tiny{HO}}}+ 2 \qquad N=-1,0,1,2,3,...\\
 &=&  \frac{1}{3}(N+1)(N+2)(N+3)+ 2 \\
& & 2,4,10,22,42,72,114,170,242,332,442,...
\end{eqnarray} 
at energies
\begin{equation} 
\label{eccr1}
E(N)_{\textrm{\tiny{CCR}}}n_{\textrm{magic 1}} = \hbar \omega_0 (N+3/2)
\end{equation} 
In addition, for $n_1 = 0$, which corresponds to the $|00 \, 2n_2 2n_2 \, 2n_3 2n_3>$ states 
and $n_2 = 0$ respectively, which corresponds to the $|2 n_1 2 n_1\, 00 \, 2n_3 2n_3>$ states 
with a
$\sum_{n=1}^{N+1} \, 2 = 2 (N+1)$-fold multiplicity for each each set $n_1 = 0$ and $n_2 = 0$ 
we obtain a second set ${_{\textrm{\tiny{CCR}}}}n_{\textrm{magic 2}} $ of  magic numbers 
\begin{eqnarray} 
\label{set2CCR}
{_{\textrm{\tiny{CCR}}}}n_{\textrm{magic 2}} &=&{_{\textrm{\tiny{CCR}}}}n_{\textrm{magic 1}}  + 4 (N+1) \qquad N=0,1,2,3,...\\
&=& \frac{1}{3}(N+1)(N+2)(N+3) +2 + 4 (N+1) \\
&=& 8,18,34,58,92,138,198,274,...
\end{eqnarray} 
at energies
\begin{equation}
\label{eccr2}
E(N)_{\textrm{\tiny{CCR}}}n_{\textrm{magic 2}} = \hbar \omega_0 (N+1)
\end{equation} 
\begin{figure}
\begin{center}
\includegraphics[width=100mm,height=144mm]{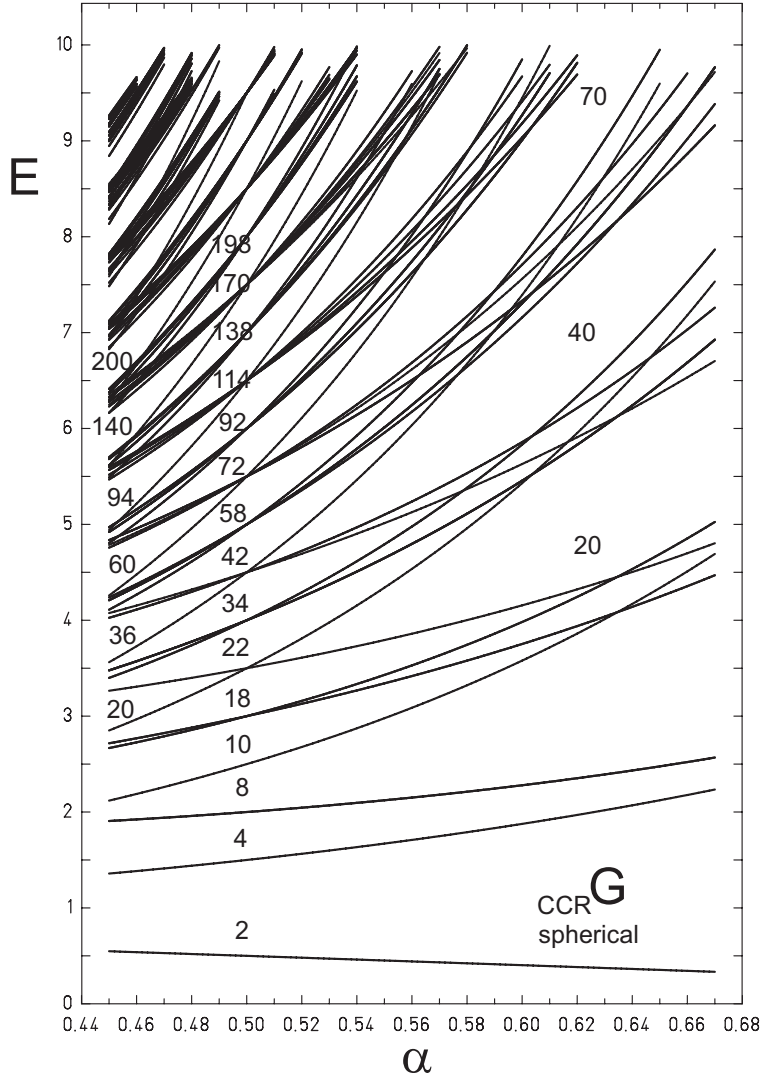}\\
\caption{\label{fig2}
level diagram of the Caputo-Caputo-Riemann symmetric rotor for the spherical case in the vicinity of
$\alpha \approx 1/2$
} 
\end{center}
\end{figure}
In figure \ref{fig2}, the single particle levels are plotted. 
A remarkable feature is the dominant influence of the $| 00 00 N N>$ state. For $\alpha< 0.5$ the harmonic oscillator
type magic numbers die out. As a consequence, for $\alpha \approx 0.48$ there remain the set of magic numbers
${_{\textrm{\tiny{CCR}}}}n_{\textrm{magic 2}}$  shifted by 1, which leads to the series $2, 4, 9,19,35,59,93,139,...$. For
$\alpha \approx 0.46$ the $| 00 00 N N>$ state has completely reached the ${_{\textrm{\tiny{CCR}}}}n_{\textrm{magic 2}}$  multiplet,
which in a stable series of magic numbers at $2,4,10,20,36,60,94,140,200,...$. 

On the other hand for $\alpha > 0.55$ the levels are rearranged to form the set of magic numbers of the harmonic oscillator.

We conclude, that the Caputo-Caputo-Riemann symmetric rotor predicts a well defined 
set of magic numbers. This set is a direct consequence of the underlying dynamic symmetries of the
three fractional rotation groups involved. It is indeed remarkable, that the same set of magic numbers 
is realized in nature as electronic magic numbers in metal clusters. 

In the next section we will demonstrate, that the proposed analytical model is an appropriate tool to 
describe the shell correction contribution to the total binding energy of metal clusters.
\section{Binding energy of metal clusters}
We will use the Caputo-Caputo-Riemann symmetric rotor (\ref{e2}) as a dynamic shell model for a description
of the microscopic part of the total energy binding energy $E_{\textrm{tot}}$ of the metal cluster.
\begin{eqnarray}
E_{\textrm{tot}} &=& E_{\textrm{macroscopic}}+E_{\textrm{microscopic}}\\
                 &=& E_{\textrm{macroscopic}}+\delta U 
\end{eqnarray}
where $\delta U$ denotes the shell-correction contributions. 

\begin{figure}
\begin{center}
\includegraphics[width=120mm,height=85mm]{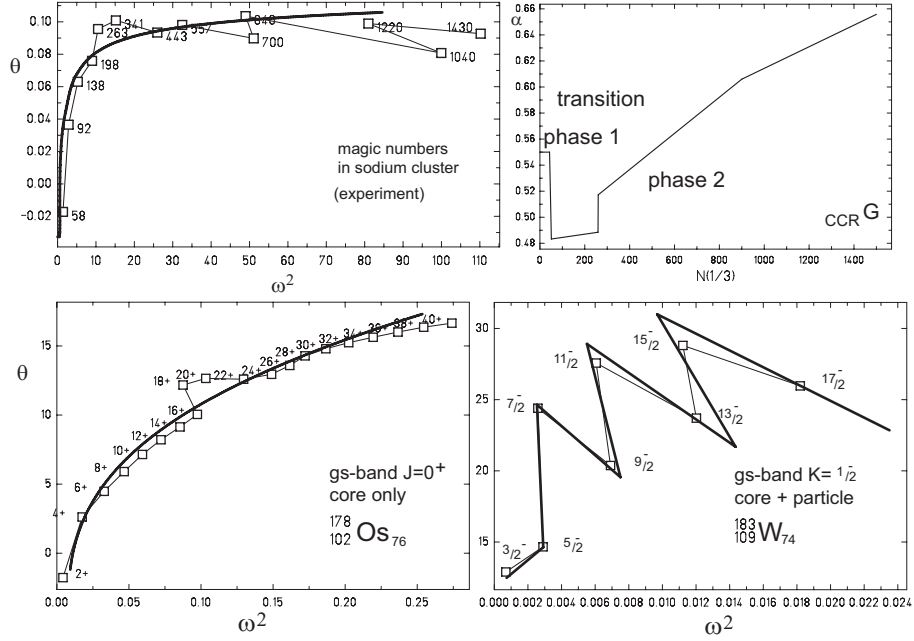}\\
\caption{\label{fig3}
Backbending plots of experimentally determined  magic numbers for (Na)$_N$ clusters from {\cite{mar}}, ground state
excitation spectrum for $^{178}_{102} \textrm{Os}_{76}$ and ground state excitation 
spectrum of $^{183}_{109} \textrm{W}_{74}$. 
Squares indicate the experimental values. 
The sequence of experimentally observed magic numbers may be categorized for $N<200$ to be 
equivalent to an excitation spectrum
of purely rotational type, the thick line indicating a fit with a fractional rigid rotor spectrum, 
for $200 \leq N \leq 300$ 
a region of backbending type (compare with the plotted ground state band spectrum for $^{178}_{102} \textrm{Os}_{76}$) 
and finally, for $N > 300$ a region with almost constant $\theta$ and a behaviour similar to excitation spectra of
ug-nuclei (compare with the plotted excitation spectrum of  $^{183}_{109} \textrm{W}_{74}$). In the upper right the
corresponding proposed $\alpha(N)$ from  (\ref{alphaN}) is plotted.   
} 
\end{center}
\end{figure}

To make our argumentation as clear as possible, we will restrict our investigation to the spherical configuration, which
will allow to discuss the main features of the proposed model in a simple context.
We will compare our results with calculations for the most prominent metal cluster, the sodium (Na) cluster. 
From experimental data\cite{kni},\cite{mar},
the following sequence of magic numbers is deduced:
\begin{eqnarray}
n_{\textrm{magic Na}} & =&  2,8,20,40,58,92,138,198 \pm 2, 263 \pm 5, 341 \pm 5, \nonumber \\ 
& &  443 \pm 5, 557 \pm 5, 700 \pm 15, 840 \pm 15, 1040 \pm 20,  \nonumber \\
& & 1220 \pm 20, 1430\pm 20
\end{eqnarray} 
For a graphical representation of the experimental magic numbers we introduce the two quantities:
\begin{eqnarray}
\Theta & =& N/(n_{\textrm{magic Na}}(N+1)-n_{\textrm{magic Na}}(N) ) \\ 
\omega & =&n_{\textrm{magic Na}}(N+1)-n_{\textrm{magic Na}}(N)  
\end{eqnarray}
Interpreting $\Theta$ as a moment of inertia and $\omega$ a rotational frequency, figure $\ref{fig3}$ 
is a backbending plot of the experimental magic numbers.  
We distinguish three different regions of magic numbers. For $N<200$ the plot shows a typical rotor spectrum. In the
region $200<N<300$ a typical backbending phenomenon is observed. For illustrative purposes in figure \ref{fig3} the same
phenomenon is documented within the framework of nuclear physics 
for the ground state rotation spectrum of   $^{178}_{102} \textrm{Os}_{76}$. For $N>300$ the moment of inertia 
becomes
nearly constant and the graph may be compared with the rotational 
$K=\frac{1}{2}$ band of the ug-nucleus  $^{183}_{109} \textrm{W}_{74}$, which is a typical example of a core plus single particle
motion in nuclear physics. 

These different structures in the sequence of electric magic numbers are reflected in the choice of the fractional
derivative coefficient $\alpha$. For $N<200$, $\alpha$ shows a simple behaviour similar to the case of magic nucleon
numbers, it varies in the vicinity of $\alpha \approx 1/2$. For the special case of sodium clusters, 
the lowest four magic numbers 
are reproduced with $\alpha>1/2$, while up to $N=198$ $\alpha< 1/2$ is sufficient. Within the backbending region there is
a sudden change in $\alpha$, which we call  a fractional second order phase transition, followed by a linear increase of the $\alpha$ value for larger cluster sizes.

The resulting dependence $\alpha(N)$ is shown in figure \ref{fig3}.

In order to compare our calculated shell correction  with published results, we use 
the Strutinsky method \cite{str1},\cite{str2} with the following parameters: 
\begin{eqnarray}
\label{om0}
\hbar \omega_0 &=& 3.96 N^{-\frac{1}{3}} [\textrm{keV}]\\
\omega_1 &= & 1\\
\omega_2 &= & 1\\
\omega_3 &= &1\\
\label{alphaN}
\alpha &=& 
\cases{
0.55               & $N < 43$ \cr 
0.908 -0.000834 \, N& $N < 51$ \cr 
0.482 +0.000025 \, N& $N< 260$ \cr
0.069 +0.000139 \, (N-260)& $N< 900$ \cr
0.062 +0.000083 \, (N-260)& $N \ge 900$ 
}\\
\label{gam}
\gamma &=& 1.1 \, \hbar \omega_0 \bigl( {_\textrm{\tiny{R}}}[N^{1/3}+1]-{_\textrm{\tiny{R}}}[N^{1/3}]\bigr)^3 \\
\label{m}
m&=& 4
\end{eqnarray}
(\ref{gam}) follows from the plateau condition $\partial U / \partial \gamma = 0$ 
and (\ref{m})
is the order of included Hermite polynomials for the Strutinsky shell correction method. 

\begin{figure}
\begin{center}
\includegraphics[width=120mm,height=76mm]{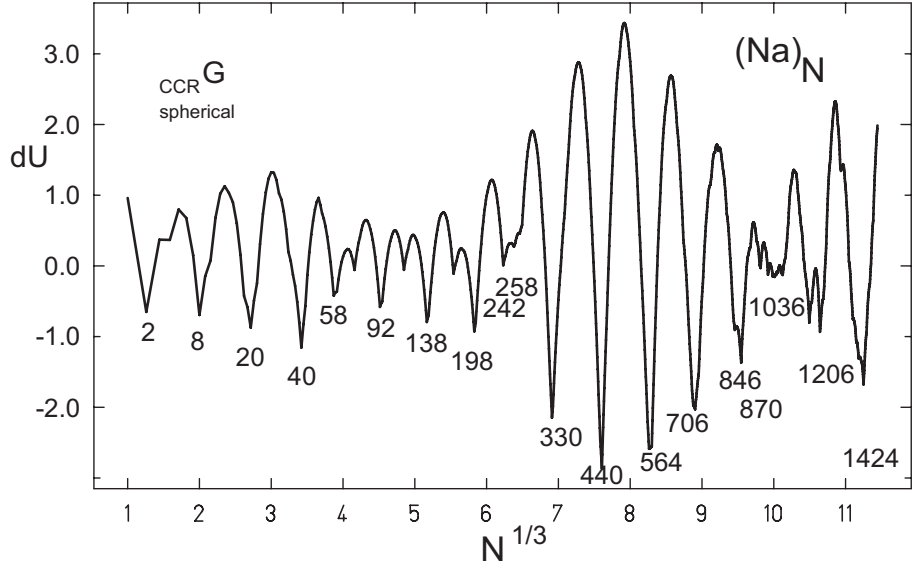}\\
\caption{\label{someshell}
Calculated shell correction $dU$ for the Caputo-Caputo-Riemann fractional symmetric 
spherical rotor with parameter set (\ref{om0})-(\ref{m})
} 
\end{center}
\end{figure}
In figure \ref{someshell} the resulting shell correction $\delta U$ is plotted. Magic numbers are reproduced correctly
within the experimental errors.
Furthermore we obtain a nearly quantitative agreement with published results for the shell correction term
obtained e.g. with the spherical Woods-Saxon potential\cite{bjo}-\cite{nis}.  

Summarizing the results presented so far, the proposed 
Caputo-Caputo-Riemann symmetric rotor describes the 
magic numbers and microscopic part of the total binding energy for 
metal clusters with reasonable accuracy. We  have demonstrated, that the cluster shell structure may
indeed be successfully described on the basis of a dynamical symmetry model.
\section{Magic numbers for clusters bound by weak and gravitational forces}
We have demonstrated, that the Caputo-Caputo-Rie\-mann rotor correctly determines  the magic numbers
of metal clusters  and that the Caputo-Riemann-Riemann rotor is an appropriate tool to describe the
ground state properties of nuclei with reasonable accuracy. Both models differ only in the mixing
ratio of fractional derivatives. The phenomena described, differ only in the interaction type of the
constituents which build up the cluster. The behaviour of metallic clusters is dominated by 
electro-magnetic forces, while in nuclei the long range part of strong forces is important.

Therefore we postulate, that the group decomposition  
\begin{equation}
{_\textrm{\tiny{RRR}}}G \supset {_\textrm{\tiny{R}}}SO^\alpha(3)
                        \supset {_\textrm{\tiny{R}}}SO^\alpha(3)
                        \supset {_\textrm{\tiny{R}}}SO^\alpha(3)\\
\end{equation}
will determine the magic number of a cluster, which is dominated by a gravitational type of interaction 
between its constituents.

The Hamiltonian ${_\textrm{\tiny{RRR}}} H$
\begin{equation}
\label{hamiltonRRR}
{_\textrm{\tiny{RRR}}} H = \frac{\omega_1}{\hbar} {_\textrm{\tiny{R}}} L_1^2(\alpha)+
    \frac{\omega_2}{\hbar} {_\textrm{\tiny{R}}} L_2^2(\alpha)+
    \frac{\omega_3}{\hbar} {_\textrm{\tiny{R}}} L_3^2(\alpha) 
\end{equation}
with the free deformation parameters $\omega_1,\omega_2,\omega_3$ on a  basis $|L_1 M_1 L_2 M_2 L_3 M_3>$
may be diagonalized and with the symmetries (\ref{sym1}) and (\ref{sym2}) a levelspectrum 
\begin{eqnarray}
\label{e1RRR}
E(\alpha) &=& \hbar \omega_1 \, {_\textrm{\tiny{R}}} [2n_1]{_\textrm{\tiny{R}}}[2n_1+1]+ 
      \hbar \omega_2 \, {_\textrm{\tiny{R}}} [2n_2]{_\textrm{\tiny{R}}}[2n_2+1]+\nonumber \\
& & 
      \hbar \omega_3 \, {_\textrm{\tiny{R}}} [2n_3]{_\textrm{\tiny{R}}}[2n_3+1] \\
\label{e2}
&=& \sum_{i=1}^3 \hbar \omega_i \frac{\Gamma(1 + (2 n_i +1) \alpha)}{\Gamma(1 + (2 n_i-1) \alpha)}  \\
& &  \qquad\qquad\qquad\qquad\qquad n_1,n_2,n_3=0,1,2,.. \nonumber
\end{eqnarray}
on a basis $| 2 n_1 2 n_1 2 n_2 2 n_2 2 n_3 2 n_3>$ results.

For the idealized spherical case $\alpha = 1/2$ this spectrum is simply given by: 
\begin{equation}
\label{eRRR}
E(\alpha=1/2) = \hbar \omega_0 (  n_1 +n_2 +n_3 +\frac{3}{2}   )
\end{equation}
which is the spectrum of the deformed harmonic oscillator. In the spherical case, magic numbers are determined
by:
\begin{eqnarray} 
\label{nRRR}
n_{\textrm{\tiny{RRR}}} &=& \frac{1}{3}(N+1)(N+2)(N+3) \qquad N=0,1,2,3,...\\
&=& 2,8,20,40,70,112,168,240,...
\end{eqnarray} 
at energies
\begin{equation} 
\label{efirst}
E(N)_{\textrm{\tiny{RRR}}} = \hbar \omega_0 (N+3/2)
\end{equation} 
This result may be compared with solutions for an independent particle shell model, where the potential
is determined by a uniformly distributed gravitational charge (mass) distribution $\rho(r) = q/V$ inside a sphere. This
potential is given by
\begin{eqnarray} 
\label{sphere}
V(r) &=& \int\int\int {\rho(r`) \over |r - r`|} d^3r` \\
     &=& q (  \frac{r^2}{2 R_0^3} - \frac{3}{2 R_0}) \qquad r<R_0 
\end{eqnarray} 
and leads to a radial Schr\"odinger equation for the harmonic oscillator. 

Therefore we are led to the conclustion, that for microscopic clusters with gravitational type of interaction
of the constituents there will be variations in the binding energy per mass unit according to (\ref{nRRR}).

Consequently we are left with the fourth decomposition of the 9-dimensional fractional rotation group
\begin{equation}
{_\textrm{\tiny{CCC}}}G \supset {_\textrm{\tiny{C}}}SO^\alpha(3)
                        \supset {_\textrm{\tiny{C}}}SO^\alpha(3)
                        \supset {_\textrm{\tiny{C}}}SO^\alpha(3)\\
\end{equation}
The Hamiltonian ${_\textrm{\tiny{CCC}}} H$
\begin{equation}
\label{hamiltonCCC}
{_\textrm{\tiny{CCC}}} H = \frac{\omega_1}{\hbar} {_\textrm{\tiny{C}}} L_1^2(\alpha)+
    \frac{\omega_2}{\hbar} {_\textrm{\tiny{C}}} L_2^2(\alpha)+
    \frac{\omega_3}{\hbar} {_\textrm{\tiny{C}}} L_3^2(\alpha) 
\end{equation}
with the free deformation parameters $\omega_1,\omega_2,\omega_3$ on a  basis $|L_1 M_1 L_2 M_2 L_3 M_3>$
may be diagonalized and with the symmetries (\ref{sym1}) and (\ref{sym2}) a levelspectrum 
\begin{eqnarray}
\label{e1CCC}
E(\alpha) &=& \hbar \omega_1 \, {_\textrm{\tiny{C}}} [2n_1]{_\textrm{\tiny{C}}}[2n_1+1]+ 
      \hbar \omega_2 \, {_\textrm{\tiny{C}}} [2n_2]{_\textrm{\tiny{C}}}[2n_2+1]+\nonumber \\
& & 
      \hbar \omega_3 \, {_\textrm{\tiny{C}}} [2n_3]{_\textrm{\tiny{C}}}[2n_3+1] \\
\label{e2}
&=& \sum_{i=1}^3 \hbar \omega_i \frac{\Gamma(1 + (2 n_i +1) \alpha)}{\Gamma(1 + (2 n_i-1) \alpha)} \nonumber \\
& & 
- \delta_{n_1 0} \hbar \omega_1 \frac{\Gamma(1 +  \alpha)}{\Gamma(1 - \alpha)}
- \delta_{n_2 0} \hbar \omega_2 \frac{\Gamma(1 +  \alpha)}{\Gamma(1 - \alpha)}
- \delta_{n_3 0} \hbar \omega_3 \frac{\Gamma(1 +  \alpha)}{\Gamma(1 - \alpha)}\\
& &  \qquad\qquad\qquad\qquad\qquad n_1,n_2,n_3=0,1,2,.. \nonumber
\end{eqnarray}
on a basis $| 2 n_1 2 n_1 2 n_2 2 n_2 2 n_3 2 n_3>$.
results. We call this model the Caputo-Caputo-Caputo symmetric rotor. For the idealized spherical case $\alpha=1/2$
this spectrum is simply given by:
\begin{equation}
\label{e12}
E(\alpha=1/2) = \hbar \omega_0 (  n_1 +n_2 +n_3 +\frac{3}{2} - \frac{1}{2}\delta_{n_1 0}- \frac{1}{2}\delta_{n_2 0}
- \frac{1}{2}\delta_{n_3 0})  
\end{equation}
\begin{figure}
\begin{center}
\includegraphics[width=100mm,height=144mm]{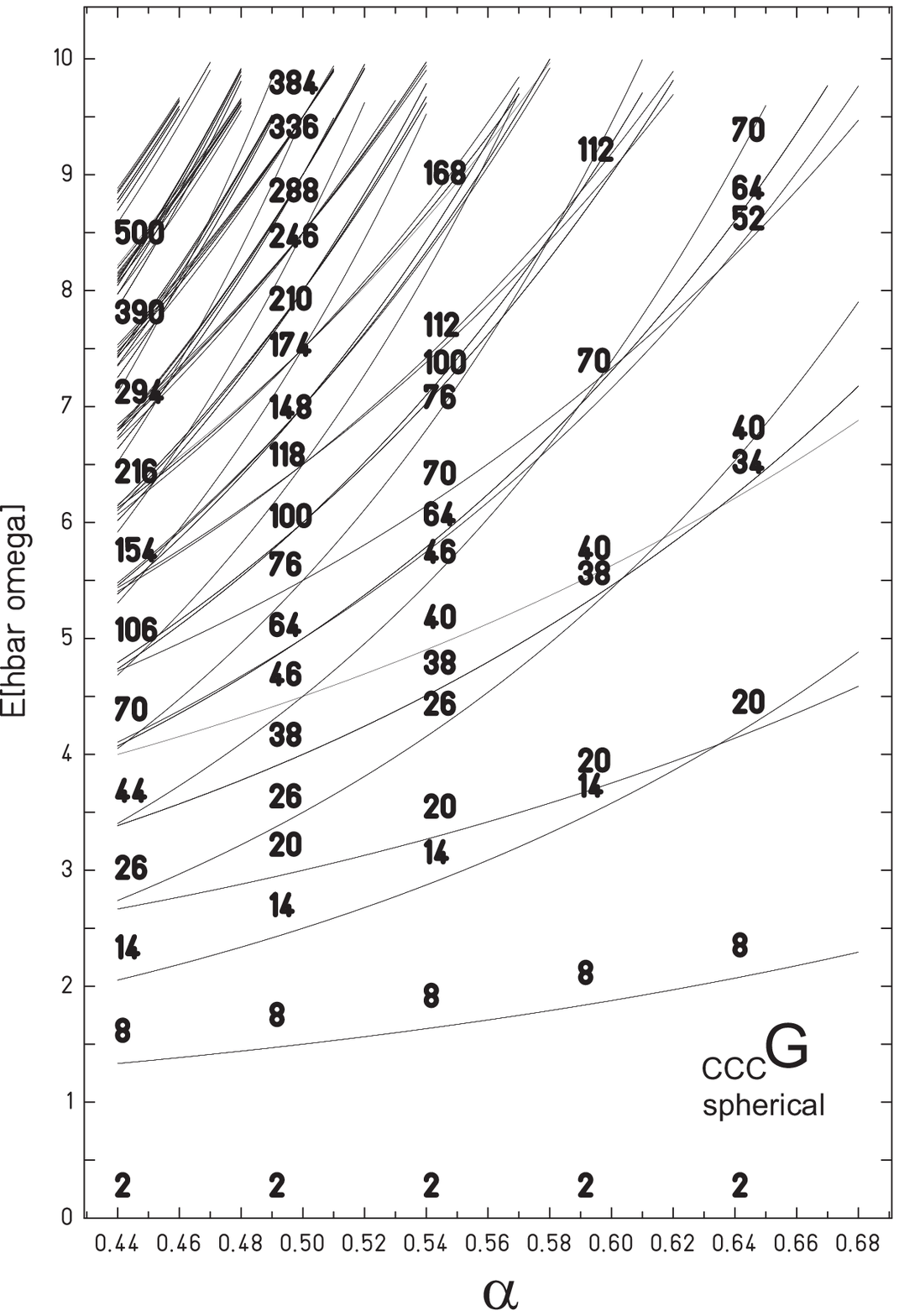}\\
\caption{\label{fig2}
level diagram of the Caputo-Caputo-Caputo symmetric rotor for the spherical case in the vicinity of
$\alpha \approx 1/2$
} 
\end{center}
\end{figure}
We obtain a first set ${_{\textrm{\tiny{CCC}}}}n_{\textrm{magic 1}} $ of  magic numbers 
\begin{eqnarray} 
\label{set1ccc}
{_{\textrm{\tiny{CCC}}}}n_{\textrm{magic 1}} &=&  n_{\textrm{\tiny{HO}}}+ 6 \qquad N=0,1,2,3,...\\
 &=&  \frac{1}{3}(N+1)(N+2)(N+3)+ 6 \\
& & 8,14,26,46,76,118,174,246,336,446,...
\end{eqnarray} 
at energies
\begin{equation} 
\label{eccc1}
E(N)_{\textrm{\tiny{CCC}}}n_{\textrm{magic 1}} = \hbar \omega_0 (N+3/2)
\end{equation} 
In addition we obtain a second set ${_{\textrm{\tiny{CCC}}}}n_{\textrm{magic 2}} $ of  magic numbers 
\begin{eqnarray} 
\label{set2CCC}
{_{\textrm{\tiny{CCC}}}}n_{\textrm{magic 2}} &=&{_{\textrm{\tiny{CCC}}}}n_{\textrm{magic 1}}  + 6 N \qquad N=1,2,3,...\\
&=& \frac{1}{3}(N+1)(N+2)(N+3) +6(N+1) \\
&=& 20,38,64,100,148,210,288,...
\end{eqnarray} 
at energies
\begin{equation}
\label{eccc2}
E(N)_{\textrm{\tiny{CCC}}}n_{\textrm{magic 2}} = \hbar \omega_0 (N+1)
\end{equation} 
Finally, the state $|000000>$ with a 2-fold multiplicity has energy $E=0$ and therefore does fit into
one of the two derived series.

Consequently we are led to the conclustion, that for microscopic clusters with weak interaction type
of the constituents there will be variations in the binding energy per charge unit according to (\ref{set2CCC}).

Summarizing the results presented in this section, we have associated the four different 
decompositions (\ref{g1})-(\ref{g4}) of the 9-dimensional mixed fractional rotation group with the
four fundamental types of interaction found in nature. We found common aspects determining the magic numbers 
for each group. There are always two different sets of magic numbers, one set is a shifted harmonic oscillator 
set, the other set is specific to the group considered. For the spherical and idealized case $\alpha=1/2$ the
four different sequences of magic number sets are simply 
the result of the presence or absence of a Kronecker-delta.

Our investigations lead to the conclusion, that mixed fractional derivative type field theories may
play an important role in a unified theory including all four fundamental interactions.					     

\section{Conclusion}
Based on the Riemann- and Caputo definition of the fractional derivative we used the fractional
extensions of the standard rotation group $SO(3)$ to construct a higher dimensional representation
of a fractional rotation group with mixed derivative types. model, which predicts the sequence of 
electronic magic numbers in metal clusters accurately. 
In the region $200<N<300$ we deduced a sudden change in the fractional derivative coefficient $\alpha$, which
we interpreted as a second order fractional phase transition.
Furthermore we have shown, that the 
microscopic part of the binding energy can be reproduced correctly  within the framework of this model.

Hence we have demonstrated, that a dynamic symmetry, generated by mixed fractional type rotation groups
is indeed realized in nature for electro-magnetic and nuclear forces. Within this framework it was possible to make
predictions for magic numbers in gravity and weak force dominated clusters.    

\section{Acknowledgment}
We thank A. Friedrich,  G. Plunien from TU Dresden/Germany  and E. Engel from  Johann-Wolfgang-Goethe-Universit\"at
Frankfurt/Germany for useful discussions.

\end{document}